\documentclass[%
aip,
amsmath,amssymb,
reprint,%
]{revtex4-1}

\usepackage{graphicx}
\usepackage{dcolumn}
\usepackage{bm}

\usepackage[utf8]{inputenc}
\usepackage[T1]{fontenc}
\usepackage{mathptmx}
\usepackage{etoolbox}

\makeatletter
\def\@email#1#2{%
	\endgroup
	\patchcmd{\titleblock@produce}
	{\frontmatter@RRAPformat}
	{\frontmatter@RRAPformat{\produce@RRAP{*#1\href{mailto:#2}{#2}}}\frontmatter@RRAPformat}
	{}{}
}%
\makeatother
\begin{document}
	
	\preprint{AIP/123-QED}
	
\title[]{Timbre Analysis of the \emph{Hulusi}, a Southwestern Chinese Free-Reed Instrument, using Machine Learning}
\author{Yang Xia}
\affiliation{College of Art, Zhejiang Normal University, Jinhua 321004, China}
\affiliation{Institute of Systematic Musicology, University of Hamburg, Hamburg 20354, Germany}
\author{Rolf Bader}
\affiliation{Institute of Systematic Musicology, University of Hamburg, Hamburg 20354, Germany}
\email{R\_Bader@t-online.de}

\date{\today} 

\begin{abstract}
The \emph{hulusi} is a wind instrument that was invented in Yunnan Province, China, and has become tremendously popular in recent years.
 It consists of a mouthpiece, a gourd, and three bamboo tubes, all with free reeds made of copper. The main bamboo tube in the middle has seven finger holes. In this instrument, the pipe length, not the free reed's eigenfrequency, determines the instrument's pitch, unlike, for example, with the Western accordion or the blues harp. In this study, a machine learning model implemented in the \emph{COMSAR} framework (https://github.com/ifsm) was used to investigate the timbre characteristics of the \emph{hulusi} to cluster different instruments and pitches. The measured \emph{hulusi} pitches C, B, A, G, and F were analyzed according to seven psychoacoustic features, among which only the spectral centroid, sharpness, and fractal correlation dimension are shown to form pitch clusters. These timbre features were used to train Kohonen self-organizing maps (SOMs) for clustering. Brightness and sharpness analysis revealed that the highest pitches were less bright and less sharp than mid- and low-range pitches were. Furthermore, the fractal correlation dimension, which mainly determines the chaoticity of the initial transients, was the best-clustering timbre feature for the \emph{hulusi}, with the highest pitches showing the least chaoticity. This result is supported by defining a cluster quality index for the SOMs.
\end{abstract}
\maketitle


\section{\label{sec:1} Introduction}
The \emph{hulusi} is a free-reed instrument that was invented in Yunnan, a Chinese province located at the southwestern border of the country with Myanmar, Laos, and Vietnam.

Wind instruments with a free reed attached to a tube with finger holes are not common in the West but are typical in Yunnan, Laos, Myanmar and Thailand.
Yunnan has the greatest density of ethnic groups worldwide and is therefore home to numerous folk songs and musical instruments. However, as Yunnan is a remote region, many of its musical instruments are not well known on a global scale. The \emph{hulusi} is an exception, as this instrument has become popular across China and the West, which might be attributed to its timbre.

The \emph{hulusi} was invented in Lianghe County on the southwest border to Yunnan. It first appeared in the 1970s, although the exact time of invention remains unknown. The instrument initially consisted of two pipes, a main pipe with fingerholes and a secondary drone pipe with only one pitch. The central pipe holes were approximately evenly spaced, and the tuning was considered inaccurate by the builders and players. Later, it evolved into a three-pipe instrument with one main pipe and two secondary drone pipes. Several \emph{hulusi} builders, including Wu, Gong, Geng \citep{Yang2017Instrument}, and others, improved the instrument in the 1980s, after which professional orchestras accepted playing the \emph{hulusi}; at this point, the instrument became popular throughout China. The \emph{hulusi} is now a mandatory instrument in primary school music education in Yunnan Province \citep{Lou2014}. Primary schools in Beijing and other cities have also implemented \emph{hulusi} teaching programs, and it has become the first instrument for many Chinese primary school students to learn music \citep{Liu2022}. Kunming, the capital of Yunnan Province, has hosted an annual \emph{Hulusi} Art Festival since 2014, featuring a unique ethnic band that plays \emph{hulusi} \citep{Liu2014}. In Lianghe County, \emph{hulusi} has become a marketing tool for tourists. In 2023, the hometown of the \emph{hulusi} welcomed 1.24 million visitors \citep{Zhang2023}.

Although the \emph{hulusi} has become popular in China, many aspects of its sound production remain unknown. It is a free-reed instrument whose sound-generating mechanism differs significantly from that of Western free-reed instruments \citep{Dieckman2006}. Similar to other Asian free-reed instruments, such as the \emph{sheng}, \emph{hulusheng}, and \emph{khaen}, the free reeds of the \emph{hulusi} are connected to a tube. However, unlike these instruments, which do not have finger holes or have only one finger hole to turn the pipe on or off, the \emph{hulusi} has seven finger holes and is therefore similar to a Western recorder.

There have been studies on an instrument similar to the \emph{hulusi}: the \emph{bawu} \citep{Cottingham2000}. The \emph{bawu} is a free-reed instrument made of a bamboo tube, with one end closed by a natural joint and the other end open. The tube body is cylindrical. A small square hole within a thin copper reed surrounded by a mouthpiece made of plastic or bone is placed near the closed end. The reed is shaped like a long and narrow isosceles triangle, which slightly bends outward when at rest.

Similar to the \emph{hulusi}, the \emph{bawu} has tone holes and a single reed and can be played by blowing into or out of the instrument. With both instruments, the reed is connected to a resonator which is typically a long round tube with a constant diameter. Therefore, the tube's sounding pitches typically have fundamental frequencies much higher than the free reed's fundamental frequency \citep{Braasch}. These frequency differences seem to result in significant variations in the sound quality of the played pitches from low to high tones, which is a crucial sound characteristic of the \emph{hulusi} and the \emph{bawu}. The effective length of the tube can be reduced by opening the sound holes during a performance, which increases the pitch. Combining reeds and tubes typically results in a sound range of less than one octave. Each tube's effective length is adjusted by cutting a large tuning slot, typically located on the tube's hidden side. The most extended pipe does not always produce the lowest pitch \citep{Cottingham2011}. However, existing studies have focused on the \emph{bawu}, \emph{sheng (sho)}, and \emph{khaen}, but research related to the \emph{ hulusi} is lacking.

All \emph{hulusi} have a key in which they are tuned. Typical keys are F, G, A, $\flat$B, or C. The fundamental pitch of the key sounds with the upper fingerholes are closed, starting from the forth hole, and the lowest three fingerholes open. We refer to this fingerhole as 4 (do) below. In the follow we name the \emph{hulusi} instruments used in the investigation by its key with an additional number in cases where more than one instrument of the same key were used. So e.g. we investigated three \emph{hulusi} in key F named F1 to F3. So the number does not refer to a register but is the index of an instrument in a certain key. 

The different timbre features of musical instruments include physical (temporal \& spectral) and psychoacoustic features. With respect to the temporal and spectral features, although the attack, decay, sustain, and release (ADSR) envelope difference and the zero-crossing rate of the temporal features are beneficial to understanding timbre, the spectral features \citep{Mcadams2013}, including the spectral centroid, spectral spread, spectral flux, and mel frequency cepstral coefficients (MFCC), play decisive roles. However, MFCC research has been criticized \citep{Alluri2010} because these features do not correspond to perceptual features; thus, three spectral features, namely, the spectral centroid, spectral spread, and spectral flux, have been emphasized in prior work. Also, loudness, roughness, and sharpness are important perceptual features. We defined loudness using the sound pressure level, which is the most commonly used indicator of sound intensity. Additionally, the fractal correlation dimension accurately describes the instrument's transient behavior and the number of harmonic overtones and is included in the present study, too.

Because the \emph{hulusi} is built differently than the instruments mentioned above, the timbre of the seven-hole \emph{hulusi} also differs. This can be specified and investigated using the psychoacoustical timbre features discussed above, such as spectral centroid, spectral flux, roughness, sharpness, or fractal correlation dimension features. To compare the timbre characteristics of different \emph{hulusis} within their pitch and timbre ranges, a music information retrieval (MIR) framework is needed that allows for the analysis and clustering of timbre characteristics, such as the librosa framework \citep{Mcfee2015}. However, the physics of reed instruments is nonlinear with respect to reed-tube coupling and wind turbulence \citep{Fletcher1978}. In the proposed models, the reed is viewed as a damped harmonic oscillator that is coupled to the pipe in a nonlinear manner, which clearly appears in features such as sudden pitch changes from the reed's eigenfrequency to that of the tube when the blowing pressure increases. This nonlinearity also shapes the initial transient of the instrument. Therefore, when MIR analysis is used, a parameter describing the nonlinearity is also needed. Here, the fractal correlation dimension in the \emph{COMSAR (https://github.com/ifsm/comsar)} sound library was chosen to measure the instrument's initial transient state, which helps in the analysis of the instrument's nonlinear behavior \citep{Bader1}. Additionally, a Kohonen self-organizing map (SOM) is used as a machine learning tool in \emph{COMSAR}. SOM is an artificial neural network that maps a high-dimensional feature space onto a two-dimensional map \citep{Kohonen1997}. Still, due to relations between neighbours in this two-dimensional representation, using SOM not as data-reduction system but as a clustering tool is straightforward in used in the present study. Unlike other artificial neural networks, e.g., deep neural networks, which are black boxes, with the SOM architecture, the reasons for successful clustering are displayed on a 2D SOM of the trained timbre feature vectors. Furthermore, the reason for successful clustering of all timbre parameters is intuitively presented using component plane visualization \citep{Martinez2025}. Therefore, in this paper, the \emph{COMSAR} framework is used to cluster the timbre characteristics of the \emph{hulusi}.

These characteristics might be caused by the frequency relations of the tube frequency at the played pitch and the eigenfrequency of the reed. This assumption is based on playing and listening experiences and the abovementioned literature. Therefore, we assume that single notes of several \emph{hulusi} will be similar and therefore cluster according to their timbre, which is defined according to psychoacoustic parameters, analogous to the timbre perception of listeners.

 Furthermore, \emph{hulusis} exist in different keys or modes, like C, F, or G key. Therefore, the same sounding pitch might be the fundamental pitch of one \emph{hulusi} while it might be a fifth on another one. Then, the relation between reed eigenmode and played pitch is very differs between instrumentso´of different modes. Therefore, we also investigate if timbre clusters remain consistent across instruments with different keys or modes. The psychoacoustic parameter clustering approach is expected to characterize the unique timbre of the instrument. In future studies, these features could be related to listeners' preferences, which could help us understand the popularity of this instrument. Measuring the reed vibrations, blowing pressures, or other physical properties is beyond the scope of the present paper and discussed in an upcoming publication.

\section{Method}
In the present study, we utilized the \emph{hulusi} made by traditional Chinese instrument makers in Yunnan Province. All the \emph{hulusis} were collected by the author in Yunnan Province during fieldwork in 2022. The author interviewed instrument builders on instrument manufacturing, materials, and related topics.

Nine \emph{hulusi} in various modes were employed. Feng Shaoxing, a skilled instrument maker in \emph{Lianghe} County, \emph{Dehong}, who has been creating \emph{hulusi} for more than 60 years, provided three of the \emph{hulusi} (in the keys of A, G, and C). Ni Kaihong of \emph{Lianghe} County, a young inheritor of the instrument, provided three \emph{hulusi} with jade mouthpieces (in the keys of C, $\flat$B, and F). The remaining three \emph{hulusi} (in the keys of C, $\flat$B, and F) were obtained from Feng Shaoxing's student, Du Deguang of \emph{Lianghe} County.

\subsection{\label{subsec:2:1} Instrument Construction}

\begin{figure}[t]
    \centering
    \includegraphics[height=0.56\columnwidth]{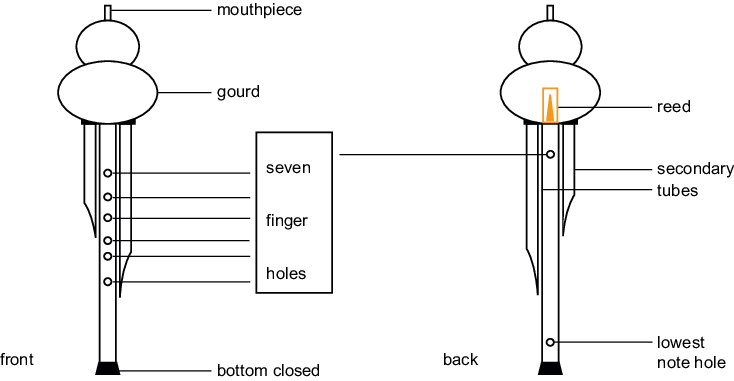}
    \caption{\emph{Hulusi} structure: front (left) and back (right) of the
    \emph{hulusi}. A player blows into the mouthpiece and the attached gourd.
    In the gourd, three tubes are placed with free reeds within the gourd.
    The middle tube has seven finger holes. The left and right tubes are
    drone tubes without finger holes and can be turned on or off using plugs
    at the tube ends.}
    \label{fig:FIG1}
\end{figure}

The modern \emph{hulusi} comprises three bamboo pipes, a mouthpiece, and a gourd. The centered primary tube is the longest component, has seven finger holes, and provides a gamut of nine full-hole pitches, encompassing one octave and a fourth. The tube is closed at the bottom. The shorter tubes on either side provide a drone; they create a sound at a fixed pitch, so no finger holes are needed. A sound is produced only when detaching plugs which are placed at the bottom of the drone tubes, whereas no sound is created when plugs are attached. Thus, the player can decide to play with or without drones. The reeds placed at the top of the three tubes are inserted into the gourd. The airflow of a player enters the gourd and then causes the reeds to vibrate. Fig.~\ref{fig:FIG1} shows the instrument's structure.

\begin{figure}[t]
    \centering
    \includegraphics[height=0.56\columnwidth]{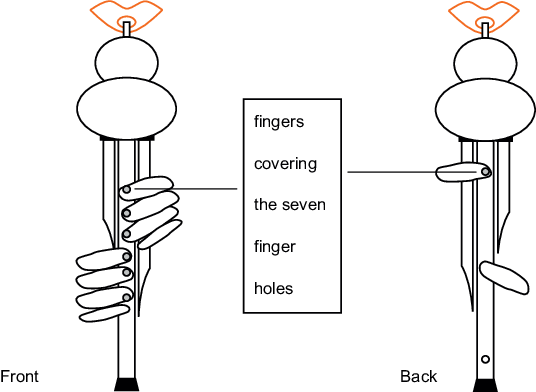}
    \caption{\emph{Hulusi} fingering for producing the lowest note:
    front (left) and back (right) of the \emph{hulusi}.}
    \label{fig:FIG2}
\end{figure}

	Fig.~\ref{fig:FIG2} shows the fingering of the \emph{hulusi} when playing the lowest note, where the left plot shows the instruments top and the right the instrument bottom, with the gray circle indicating the holes covered by the fingers. The pitch increases as the holes are opened sequentially from bottom to top, and the highest note is played by opening all the holes, with the last open hole on the back of the tube, as shown on the right side of Fig.~\ref{fig:FIG2}. Notably, when the pitch of fa is played, the top hole is opened while the bottom holes are covered, resulting in lower air pressure when playing this note compared to other notes. 
	
	Table 1 shows \emph{hulusi} in various keys, or lowest note on the instrument, from C1 to $\flat$B2. For each key all nine possible notes are displayed, using solmization do, re, mi, fa, sol la, si. Note that the keys are labeled for fingering four, where the fundamental notes of the keys are fingering 4. Also, the fundamental frequencies for each note is displayed as measured by playing the instrument in Hz, as discussed below.
	
\begin{table*}[t!]

\caption{\label{tab:hulusi_compact}
Nine \emph{hulusis} in keys F1 to C3, showing note names, solmization,
and measured fundamental frequencies of the notes played on the instruments.
Values following the note names are the mean fundamental frequencies
$\pm$ standard deviations in Hz obtained from ten measurements.
For each instrument, the second row gives the interval in cents relative
to fingering 4 (\emph{do}), which is defined as 0 cents.
}

\centering
\small

\setlength{\tabcolsep}{5.0pt}
\renewcommand{\arraystretch}{1.12}

\begin{ruledtabular}
\begin{tabular}{lccccccccc}

Instrument &
\multicolumn{9}{c}{Fingering} \\

\cline{2-10}

&
1 (usol) &
2 (ula) &
3 (usi) &
4 (do) &
5 (re) &
6 (mi) &
7 (fa) &
8 (sol) &
9 (la) \\

\hline

F1
& C4 265$\pm$1
& D4 293$\pm$1
& E4 328$\pm$1
& F4 348$\pm$1
& G4 389$\pm$1
& A4 439$\pm$1
& B$\flat$4 468$\pm$2
& C5 521$\pm$0
& D5 584$\pm$1 \\[-1pt]

{\scriptsize\emph{cents}}
& {\scriptsize $-471.7$}
& {\scriptsize $-297.8$}
& {\scriptsize $-102.5$}
& {\scriptsize $\mathbf{0.0}$}
& {\scriptsize $192.8$}
& {\scriptsize $402.2$}
& {\scriptsize $512.9$}
& {\scriptsize $698.6$}
& {\scriptsize $896.3$}
\\[3pt]

F2
& C4 265$\pm$1
& D4 294$\pm$1
& E4 328$\pm$1
& F4 347$\pm$1
& G4 389$\pm$1
& A4 437$\pm$1
& B$\flat$4 474$\pm$1
& C5 520$\pm$2
& D5 584$\pm$1 \\[-1pt]

{\scriptsize\emph{cents}}
& {\scriptsize $-466.7$}
& {\scriptsize $-286.9$}
& {\scriptsize $-97.5$}
& {\scriptsize $\mathbf{0.0}$}
& {\scriptsize $197.8$}
& {\scriptsize $399.2$}
& {\scriptsize $539.9$}
& {\scriptsize $700.3$}
& {\scriptsize $901.2$}
\\[3pt]

G
& D4 296$\pm$1
& E4 332$\pm$1
& F$\sharp$4 371$\pm$1
& G4 394$\pm$1
& A4 442$\pm$1
& B4 497$\pm$1
& C5 530$\pm$2
& D5 593$\pm$2
& E5 666$\pm$2 \\[-1pt]

{\scriptsize\emph{cents}}
& {\scriptsize $-495.1$}
& {\scriptsize $-296.4$}
& {\scriptsize $-104.1$}
& {\scriptsize $\mathbf{0.0}$}
& {\scriptsize $199.0$}
& {\scriptsize $402.1$}
& {\scriptsize $513.4$}
& {\scriptsize $707.8$}
& {\scriptsize $908.8$}
\\[3pt]

A
& E4 332$\pm$1
& F$\sharp$4 374$\pm$1
& A$\flat$4 416$\pm$2
& A4 441$\pm$2
& B4 497$\pm$1
& C$\sharp$5 558$\pm$1
& D5 573$\pm$2
& E5 663$\pm$2
& F$\sharp$5 744$\pm$2 \\[-1pt]

{\scriptsize\emph{cents}}
& {\scriptsize $-491.5$}
& {\scriptsize $-285.3$}
& {\scriptsize $-101.0$}
& {\scriptsize $\mathbf{0.0}$}
& {\scriptsize $207.0$}
& {\scriptsize $407.4$}
& {\scriptsize $453.3$}
& {\scriptsize $705.9$}
& {\scriptsize $905.4$}
\\[3pt]

$\flat$B1
& F4 352$\pm$0
& G4 393$\pm$1
& A4 440$\pm$1
& B$\flat$4 466$\pm$1
& C5 525$\pm$2
& D5 588$\pm$0
& E$\flat$5 626$\pm$2
& F5 699$\pm$0
& G5 785$\pm$0 \\[-1pt]

{\scriptsize\emph{cents}}
& {\scriptsize $-485.7$}
& {\scriptsize $-295.0$}
& {\scriptsize $-99.4$}
& {\scriptsize $\mathbf{0.0}$}
& {\scriptsize $206.4$}
& {\scriptsize $402.6$}
& {\scriptsize $511.0$}
& {\scriptsize $702.0$}
& {\scriptsize $902.8$}
\\[3pt]

$\flat$B2
& F4 354$\pm$1
& G4 397$\pm$1
& A4 443$\pm$1
& B$\flat$4 482$\pm$1
& C5 529$\pm$1
& D5 595$\pm$0
& E$\flat$5 611$\pm$2
& F5 703$\pm$2
& G5 785$\pm$1 \\[-1pt]

{\scriptsize\emph{cents}}
& {\scriptsize $-534.3$}
& {\scriptsize $-335.9$}
& {\scriptsize $-146.1$}
& {\scriptsize $\mathbf{0.0}$}
& {\scriptsize $161.1$}
& {\scriptsize $364.6$}
& {\scriptsize $410.6$}
& {\scriptsize $653.4$}
& {\scriptsize $844.4$}
\\[3pt]

C1
& G4 398$\pm$2
& A4 440$\pm$2
& B4 489$\pm$2
& C5 520$\pm$1
& D5 581$\pm$1
& E5 652$\pm$2
& F5 703$\pm$9
& G5 775$\pm$2
& A5 871$\pm$2 \\[-1pt]

{\scriptsize\emph{cents}}
& {\scriptsize $-462.9$}
& {\scriptsize $-289.2$}
& {\scriptsize $-106.4$}
& {\scriptsize $\mathbf{0.0}$}
& {\scriptsize $192.0$}
& {\scriptsize $391.6$}
& {\scriptsize $522.0$}
& {\scriptsize $690.8$}
& {\scriptsize $893.0$}
\\[3pt]

C2
& G4 397$\pm$1
& A4 445$\pm$2
& B4 495$\pm$2
& C5 529$\pm$1
& D5 590$\pm$2
& E5 663$\pm$2
& F$\sharp$5 721$\pm$2
& G5 785$\pm$2
& A5 881$\pm$2 \\[-1pt]

{\scriptsize\emph{cents}}
& {\scriptsize $-497.0$}
& {\scriptsize $-299.4$}
& {\scriptsize $-115.0$}
& {\scriptsize $\mathbf{0.0}$}
& {\scriptsize $188.9$}
& {\scriptsize $390.9$}
& {\scriptsize $536.1$}
& {\scriptsize $683.3$}
& {\scriptsize $883.0$}
\\[3pt]

C3
& G4 395$\pm$1
& A4 441$\pm$1
& B4 493$\pm$2
& C5 520$\pm$1
& D5 583$\pm$2
& E5 656$\pm$1
& F5 694$\pm$2
& G5 780$\pm$1
& A5 875$\pm$2 \\[-1pt]

{\scriptsize\emph{cents}}
& {\scriptsize $-476.0$}
& {\scriptsize $-285.3$}
& {\scriptsize $-92.3$}
& {\scriptsize $\mathbf{0.0}$}
& {\scriptsize $198.0$}
& {\scriptsize $402.2$}
& {\scriptsize $499.7$}
& {\scriptsize $702.0$}
& {\scriptsize $900.9$}
\\

\end{tabular}
\end{ruledtabular}

\end{table*}

\subsection{\label{subsec:2:2} Measurements}

The author played and recorded single tones and melodies on nine \emph{hulusis} in the anechoic chamber at the Institute of Systematic Musicology at the University of Hamburg using a dummy head (Head Acoustics SQuadriga Type HSU III.2). Table 1 lists the different \emph{hulusi} modes ($\flat$B, C, A, G, F) collected during the fieldwork, two F-, one G- and A-, two $\flat$B-, and three C-keyed instruments.

Notes played by a musician were used to approximate the real-world performance characteristics of the instrument as closely as possible. As the instrument is played through a mouth hole into a gourd, a very standardized repetition of these notes is possible. Therefore, the timbre crucially depends on the playing pressure, which is relatively easy to keep constant. A pressure that is too low results in the instrument playing a sound with the reed-dominate-frequency and not with that of the tube-dominate-frequency. Therefore, a minimum pressure is needed to reach the tube pitch. At each initial transient, the pressure increases; therefore, for each initial transient, the reed-dominate-frequency first appears briefly. This contributes strongly to the overall sound of the instrument. In the upper limit, if the pressure is too high, the pitch increases, the instrument overblows, or the instrument does not make any sound. Therefore, larger deviations in the sound pressure used in the ten repetitions would have led to considerable differences in timbre and pitch. Therefore, a standard pressure over all the cases was easy to achieve.

Each pitch of each instrument was played ten times and recorded for three seconds at sample rate of 44.1 kHz. All 810 recordings, comprising 9 \emph{hulusi} $\times$ 9 pitches $\times$ 10 cases of each pitch, were averaged with respect to the 10 measurements of each pitch, resulting in 81 mean pitches. Then, each averaged note was split into 50 overlapping time windows of frame length of $2^{13}$ samples. All frames were Fourier analyzed to obtain a temporal development of the 81 notes.

\subsection{\label{subsec:2:3} Psychoacoustic Parameter Estimation using the MIR Framework}

Sound analysis was performed using the \emph{COMSAR} framework developed at the Institute of Systematic Musicology during the Computational Recording Archive project \citep{Bader2}. The framework is written in Python and can be run in a browser environment via a Python-based Jupyter notebook. \emph{COMSAR} includes two melody libraries, tone system libraries, and a timbre feature library. In this paper, we extract acoustic features from hulusi recordings using the timbre feature library and then map these features onto psychoacoustic metrics via established hearing models \citep{BlassWeb}. The parameters in this library include the spectral centroid, spectral spread, spectral flux, roughness, sharpness, sound pressure level, and fractal correlation dimension.

\subsubsection{Spectral Centroid}

The spectral centroid C represents the timbre brightness and is one of the primary features when the timbre dissimilarity is rated \citep{Blass2019}. It is the sum of the amplitude $A_i$ weighted frequencies $f_i$ normalized by the sum of all the amplitudes as follows:

\begin{eqnarray}
	C = \frac{\sum^{N}_{i=1}A_if_i}{\sum^{N}_{i=1}A_i} \ ,
	\label{eq:one}
\end{eqnarray}

where $ A_i $ refers to the amplitude value of the spectrum of the i-th frequency point and $ f_i $ is the frequency value corresponding to the i-th frequency of a discrete spectrum obtained from a time series.

\subsubsection{Spectral Spread}
The spectral spread indicates the degree of distribution of the frequency content in a spectrum and can be used to distinguish between noise and pitch sounds. It can be obtained by calculating the deviation between the spectrum and spectral centroid as follows:
\begin{eqnarray}
	SS = \sqrt{\frac{\sum^{N}_{i=1} \left( A_i-C \right)f_i}{\sum^{N}_{i=1}A_i}} 
	\label{eq:one}
\end{eqnarray}

\subsubsection{Spectral Flux}
The spectral flux indicates the rate at which the power in the audio spectrum changes. It is the amount of change in the spectrum between two adjacent sound analysis frames. 

Several spectral flux definitions have been proposed, and in this paper, we use the approach of scaling the spectral flux in each window by the size of the current window \citep{Blass2019}:
\begin{eqnarray}
	SF \left( t \right) = \frac{\sum^{}_{n} H \left( \vert X \left( t,n \right) \vert - \vert X \left( t,n-1 \right) \vert \right)}{\vert \vert X \left( t,n \right) \vert \vert}
	\label{eq:one}
\end{eqnarray}

Here, $ X \left(n \right) $ and $ X \left(n-1 \right) $ refer to the banded energy difference between the spectra of two continuous time windows, where n is an integer index.

\subsubsection{Roughness}
Roughness is the primary form of auditory perception caused by fast amplitude modulations \citep{Zwicker2013}. In this paper, the Helmholtz--Bader algorithm \citep{Bader1} is used to calculate roughness. When the difference in frequency between two adjacent spectral peaks is approximately 15 Hz, the sound begins to sound rougher, whereas the maximum roughness of the two sinusoidals is assumed to occur at a frequency difference of 33 Hz, as follows:

\begin{eqnarray}
	R_n = A_1 A_2 \frac{\vert {\it d}f_n \vert}{f_r {\it e}^{-1}} {\it e}^{- \vert {\it d} f_n \vert / f_r} 
	\label{eq:one}
\end{eqnarray}

Here, $ A_1 $ and $ A_2 $ are the amplitudes of the two frequencies, and the distance between them is $ {\it d} f_n $. Although this measure might be too simple for very low frequencies, for the frequency range of the \emph{hulusi}, it is assumed to be appropriate.

The roughness R then equals the sum of all possible combinations of sine waves:

\begin{eqnarray}
	R = \sum_{i=1}^n R_i
	\label{eq:one}
\end{eqnarray}

\subsubsection{Sharpness}
Sharpness measures a sound's high-frequency content. The spectrum content and center frequency of the narrowband sound are the most important parameters influencing sharpness, with the bandwidth also being an influencing factor \citep{Fastl2007}. Sharpness is measured in acum, and one acum represents narrowband noise with a level of 60 dB and a center frequency of 1 kHz in the critical band. Narrowband noise increases sharply above 3 Hz but increases only slightly below 3 Hz, as follows:

\begin{eqnarray}
	Sh = 0.11 \frac{\int_0^{24Bark}{L_B g \left( z \right) z}{\rm d}z}{\int_0^{24Bark}{L_B}{\rm d}z}acum 
	\label{eq:one}
\end{eqnarray}

Here, $ L_B $ represents the total loudness of each bark band, while $ g \left(z \right) $ is an additional weighting factor that increases from 1 to 4 at critical bands greater than 16, as follows:

\begin{eqnarray}
	g \left(z \right) = \begin{cases}
		1 & z \leq 16 \\
		0.066 e^{0.171B} & z>16 \\
	\end{cases} \ .
	\label{eq:one}
\end{eqnarray}

\subsubsection{Sound Pressure Level (SPL)}
The sound pressure level (SPL) is a common physical measure of acoustic power and is calculated as follows:

\begin{eqnarray}
	L_p = 20 \log_{10} \left({\frac{p}{p_0}} \right)dB \ ,
	\label{eq:one}
\end{eqnarray}

where $ p $ is the measured root-mean-square (RMS) sound pressure of the wave and $ p_0 $ is the reference sound pressure. In the present study, the reference value is $ 2 \times 10^{-5}$ Pa  \citep{Rossing2007}.

\subsubsection{Fractal Correlation Dimension}
The fractal correlation dimension is useful for analyzing the initial transients of musical instruments and counting the number of nonharmonic components/overtone structures and significant amplitude fluctuations in musical sounds. It can be defined by dividing the logarithm of the parameters $ C \left( r \right) $ and $ r $ \citep{Bader2013} as follows:

\begin{eqnarray}
	D = \frac{\ln C \left( r \right)}{\ln r} 
	\label{eq:one}
\end{eqnarray}

The parameter $ C \left( r \right) $ is the sum of all distances between all sampled sound amplitudes in an n-dimensional space that are less than radius $ r $ apart, whereas $ X_i $ and $ X_j $ are one-dimensional time series amplitudes $ X_t$ embedded at any two points in an n-dimensional space. $ r $ is the maximum distance in this space, and the Heaviside function $ H \left( . \right) $ is introduced by counting the number of distances within this radius as follows:

\begin{eqnarray}
	C \left( r \right) = \frac{1}{N^2} \sum_{i \neq j} H \left( r- \vert X_i - X_j \vert \right) \ ,
	\label{eq:one}
\end{eqnarray}

where

\begin{eqnarray}
	H \left( x \right) = \begin{cases}
		0 & for\ x\ \leq 0 \\
		1 & for\ x\ >0 \\
	\end{cases}
	\label{eq:one}
\end{eqnarray}

For the 81 averaged pitches recorded on the nine \emph{hulusi}, as discussed above, these psychoacoustic features were calculated. Therefore, for each note a  psychoacoustic feature vector exists with seven entries, according to the seven psychoacoustic features. These features are fed into the SOM in various combinations, discussed below.

\subsection{\label{subsec:2:4} Machine Learning}
In this paper, the \emph{COMSAR} architecture developed at the Institute of Systematic Musicology, which employs SOMs, is used. For the trained 2D map, a neural grid of 18$\times$18 neurons was chosen. Each neuron is a feature vector with a certain number of psychoacoustic feature entries chosen to analyze each of the 81 cases. The SOM is then trained in two steps. First, all the neurons are randomly initiated with normalized vectors. Second, each of the 81 training vectors is correlated with all 18$\times$18 map vectors, and the map vector with the highest correlation with the training vector is found. This map vector and its surrounding area are altered on the basis of the training vector. The surrounding area is altered according to a Mexican hat function. This correlation and alternation is performed for each of the 81 training vectors, resulting in one training iteration. Five hundred training iterations were performed for the SOM to converge according to differences in the training feature vector set.

The training results are displayed as a Kohonen \emph{SOM} map, visualizing the so-called u-matrix. The U-matrix (unified distance matrix) is a representation of a self-organizing map (SOM) where the Euclidean distance between the codebook vectors of neighboring neurons is depicted in a grayscale image. The u-matrix presents the mean correlations of one trained SOM vector with its neighboring vectors. During training, the SOM iteratively adjusts the codebook vectors of the 2‑dimensional output nodes to preserve the topological structure of the high‑dimensional input space. As a result, regions of similar trained vectors emerge on the map, separated by boundaries where neighboring neurons differ significantly. This leads to map regions with highly consistent trained vectors and regions where neighboring neurons are considerably different.

\begin{figure}
	\includegraphics[scale=1]{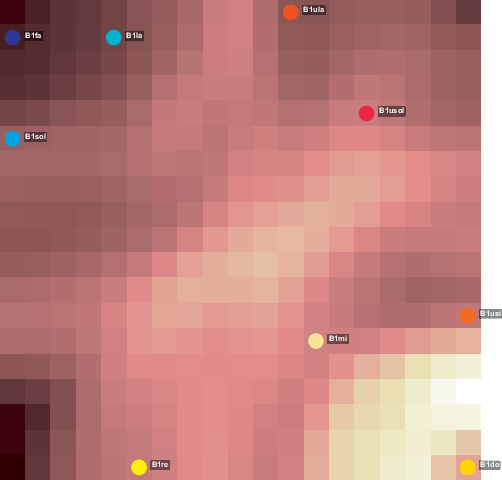}
	\caption{\label{fig:FIG3}{The results of training a self-organizing map of the \emph{hulusi's } timbre features using an 18 $\times$ 18 u-matrix. The notes used to train the \emph{SOM} are represented by colored dots, with different colors representing the instrument's pitches: the blue dots in the lower left corner represent the treble region, the orange dots represent the midrange region, and the pink dots represent the bass region.}}
	\raggedright
\end{figure}

Regions of high consistency often contain clusters and have boundary ridges with low correlation values, whereas other regions display high internal consistency.
 In the SOM, darker colors indicate regions of high consistency, whereas light colors indicate substantial differences between trained neighboring neurons. Thus, light ridges represent boundaries between consistent regions. As shown in each map in Fig.~\ref{fig:FIG3}, the dark colors indicate small distances between neighboring cells, i.e., clustering regions, whereas the light colors indicate considerable distances between neighboring cells, i.e., clustering boundaries.

The pitch timbre feature vectors are then fitted to the trained SOM and placed at the corresponding best-fit positions. The best fit is determined on the basis of the correlations between a training vector and all trained SOM neurons, and the neuron with the highest correlation is selected. The color and text next to the best-fit pitches on the map indicate the sounding pitch of the \emph{hulusi} note. Therefore, the map allows us to examine the clustering of pitches according to the timbre feature vector based on which the map was trained.

Fig.~\ref{fig:FIG3} shows an example of a Kohonen SOM obtained through single-parameter training, with the spectral centroid shown as an example. The background shows the similarity/difference of the spectral centroid features of the trained SOM. The dark background represents clustering areas, whereas the light background represents areas with large spectral centroid differences.

According to the SOM analysis, the clustering results might improve as the number of features used increases but decrease when too many features are used in the training vector.
This might correspond to timbre perception investigations using multidimensional scaling methods (MDSs), where rarely more than three timbre features were used by listeners to discriminate musical instrument timbres \citep{Bader2013}.

\section{Results}
\label{results}

\subsection{\label{subsec:3:1}Single-Feature SOMs}
As mentioned above, the number of features used in the SOM are varied. To obtain a systematic understanding of the impact of features to feature combinations, single features are initially used to create the SOMs. Therefore, at first, for the whole set of 81 averaged \emph{hulusi} notes, seven SOMs were calculated, one for each psychoacoustic feature. Only those features for which clustering is observed are selected for feature combinations. Additionally, the ability of each feature to cluster the timbres is estimated.

\begin{figure*}
\includegraphics[scale=1]{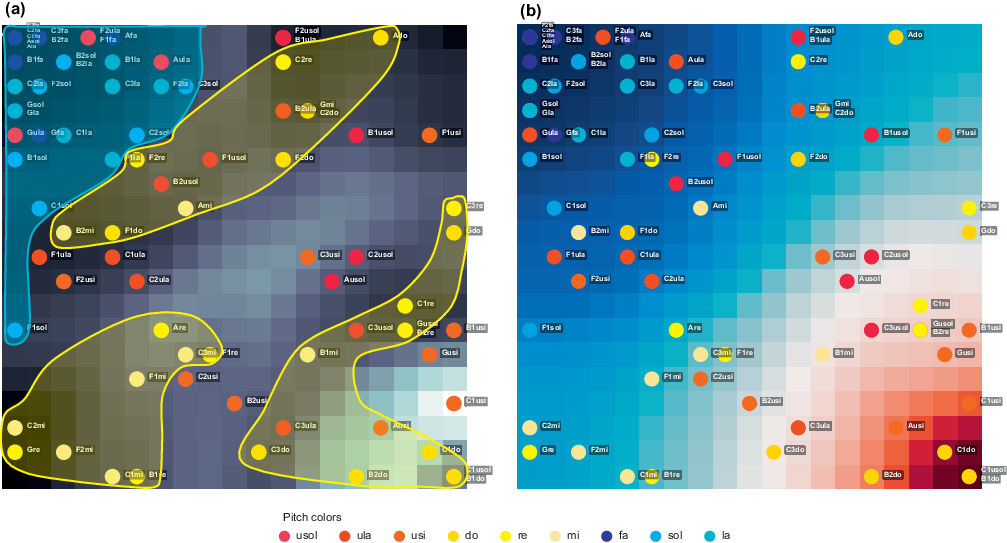}
\caption{\label{fig:FIG4}SOM of the spectral centroid. Left: U-matrix, the background color represents the similarity of neighboring neurons, darker regions represent lower values. Right: component plane; the background color represents the spectral centroid feature. The dot colors represent the recorded pitches. The pitches are arranged from high to low. Treble tones: blue, where \emph{la, sol,} and \emph{fa} transition from light blue to dark blue. Midrange tones: yellow, where mi, re, and do transition from light yellow to yellow. Bass tones: red, where \emph{usi, ula,} and \emph{usol} transition from orange to dark red.
}
\vskip1in
\end{figure*}

Fig. 4 and Fig. 5 depict the single-feature SOMs for four timbre features: the spectral centroid, sharpness, fractal correlation dimension, and spectral flux. After training, the sounds used for training were fitted to the best-matching neuron in the trained map, as indicated by the dots. In each plot, pitches are represented by dot color, ranging from G3 (dark red) to A4 (light blue), a span of one octave and one major second (a major ninth). Different parameters were used to determine the background colors for the left and right plots in Fig.~\ref{fig:FIG4}. The background color in the left plot represents the similarity of the neighboring neurons, with dark-colored regions indicating higher similarity and clustering regions. The light colors represent regions of lower similarity between neighboring neurons. Therefore, ridges of light color serve as cluster boundaries. The background color of the right panel corresponds to the trained features and is a feature or component map. The background color of the right plot in Fig.4 indicates the spectral centroid feature. The sharpness, fractal correlation dimension, and spectral flux features are shown in Figs. 5(b), (d), and (f), respectively. In these feature maps, dark colors represent low similarity, and light colors represent high similarity.

The left plot in Fig.~\ref{fig:FIG4} shows that the treble tones \emph{sol, la,} and \emph{fa} are regularly clustered in the upper left region. The middle sounds \emph{mi, re,} and \emph{do} are scattered on both sides of the light-colored border in the middle region, as shown by the yellow color in the figure. The do pitches cluster in the bottom right corner: the bass tones \emph{usi, ula,} and \emph{usol} are dispersed throughout and do not form a cluster.

The right plot in Fig.~\ref{fig:FIG4} depicts the same \emph{SOM}, now showing the spectral centroid as background color. The dark and light colors denote low and high spectral centroid values, respectively. A continuous background color indicates a gradient from low values (lower right corner) to high values (upper left corner). The u-matrix on the left side of Fig. 4 contains ridges and is discontinuous, indicating that the gradient is not continuous and instead changes. Therefore, a 2D representation of a one-dimensional feature (e.g. the spectral centroid) can be obtained, as 2D mapping allows for differentiation of this discontinuity.

Interestingly, the centroid of the mid-range tones is higher than that of the high-range tones, which is counterintuitive. Indeed, the high-pitched sounds have the smallest brightness values, and even most low tones are brighter. This is caused by the high tones being `thin', i.e., having only a few overtones compared with the mid- and low-range pitches.

Also note that the bass notes are spread throughout the plot and therefore do not follow a simple continuous relation between note and brightness. 

These two findings indicate that the spectral centroid features of the midrange and high-range pitches are continuous. In contrast, the spectral centroid features of the low-range pitches are scattered, indicating unsystematic brightness variations in the \emph{hulusi} bass tones. A closer examination reveals the high-pitched notes in a single cluster in the left upper corner, whereas the midrange notes form three distinct clusters.

\begin{figure*}
\includegraphics[scale=0.9]{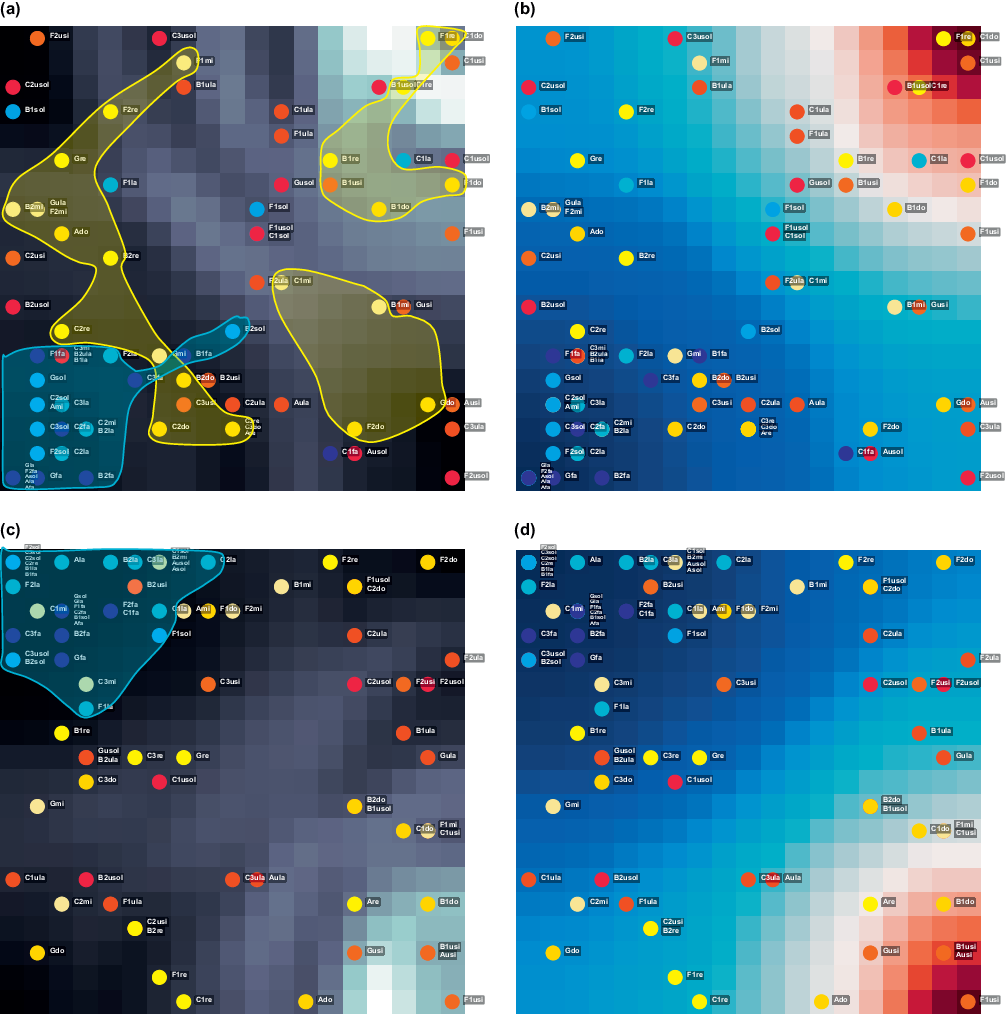}\\
\includegraphics[scale=0.9]{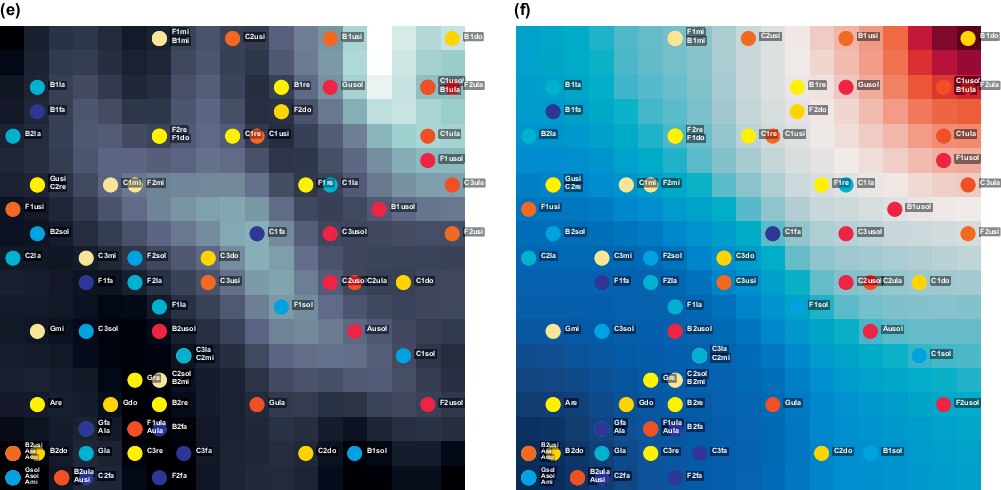}
\caption{\label{fig:FIG5}SOMs and component maps of the sharpness (a, b), fractal correlation dimension (c, d), and spectral flux (e, f) features. The background color corresponding to neighboring neuron similarity is shown on the left, and the background color corresponding to the feature strength is shown on the right. The sharpness and fractal dimension plots show clusters for high-pitched tones but not for midrange and low-range registers. The flux results do not display any clustering, and the feature strength does not align with the pitches.
}
\end{figure*}

Figs.~\ref{fig:FIG5}(a) and (b) show the sharpness results, with (a) displaying the neuron similarity and (b) showing the feature map as the background color. The dot distributions display clustering regions and cluster boundaries.

The treble pitches \emph{sol, la,} and \emph{fa} are clustered in the lower left area. The midrange sounds \emph{mi, re,} and \emph{do} are primarily clustered in the middle areas (slightly lighter colored region), with a few pitches in the upper right corner in separate regions distinguished by the clustering boundaries, as shown by the yellow curve region in the figure.
Similar to the spectral centroid feature, the bass pitches \emph{usi, ula,} and \emph{usol} are dispersed and not clustered in the sharpness SOM. Thus, the sharpness feature can be used to clearly distinguish high-, mid-, and low-range notes, with the midrange clustering area distributed on both sides of the light color boundaries.

The sharpness feature is used to provide the background color in Fig.~\ref{fig:FIG5}(b). The midrange sounds have higher overall sharpness values than the treble sounds. Similarly, the midrange sounds have more overtones and thus more sounds at higher frequencies than the treble sounds, which is consistent with the brightness feature results and again a counterintuitive finding.

Fig.~\ref{fig:FIG5}(c) and (d)  show the fractal correlation dimension results, which are slightly different from those of the previous two features; in the plot of the similarity background in (c), the treble \emph{sol, la,} and \emph{fa} sounds are perfectly clustered in the darker-colored region in the upper left corner. As the pitch decreases, the alto \emph{mi, re,} and \emph{do} sounds gradually move not only to the lighter-colored clustered boundary region in the upper right region but also to other regions outside the upper left corner. The bass \emph{usi} and \emph{ula} sounds are located in the lower right region, while the \emph{usol} sound is found in the upper left corner. Thus, the data reveal that only the treble \emph{sol, la,} and \emph{fa} sounds form a clear cluster.

As shown in Fig.~\ref{fig:FIG5}(d), in which the fractal dimension feature is the background, alto sounds are relatively dispersed throughout the map but have higher correlations and are therefore more chaotic than the treble sounds. The correlation is proportional to the number of inharmonic overtones, especially during the initial transient. Therefore, the timbre of the low- and mid-range pitches is more complex than that of the higher pitches, which is expected from and consistent with the findings above.

The spectral flux characteristics in Figs.~\ref{fig:FIG5}(e) and (f) show very different scenarios. As shown in (e), most treble and midrange pitches are clustered in a darker region in the lower left with a similar background color. A few treble pitches are clustered in a slightly brighter area in the lower right. A brighter boundary forms along the diagonal in the center, with a small cluster of bass and midrange sounds in the upper right region and a cluster of bass pitches in the upper left. These results show that it is impossible to cluster different pitches on the basis of flux characteristics, and only rough estimates can be made. The background plot of the spectral flux features in Fig.~\ref{fig:FIG5}(f) shows that the midrange and treble tones are mixed, indicating that there is no significant difference between their spectral fluxes.

The results for the remaining three timbre features, namely, the spectral spread, SPL, and roughness, are not displayed here as, similar to the spectral flux, no clusters appear with these features. These features are discussed statistically in Section \ref{subsec:3:3} below.

\subsection{\label{subsec:3:2} Multifeature SOMs}
All the results thus far are clustered using single features to show the feature clustering capabilities of the proposed approach. By combining single features with the feature vectors, improved clustering can be achieved.

As shown above, the spectral centroid and sharpness cluster \emph{hulusi} treble and midrange tones well, so all of the feature combinations used below include these two features.

\begin{figure*}[t]
\centering

\begin{minipage}[t]{0.48\textwidth}
\vspace{0pt}
\centering
\includegraphics[width=1\linewidth]{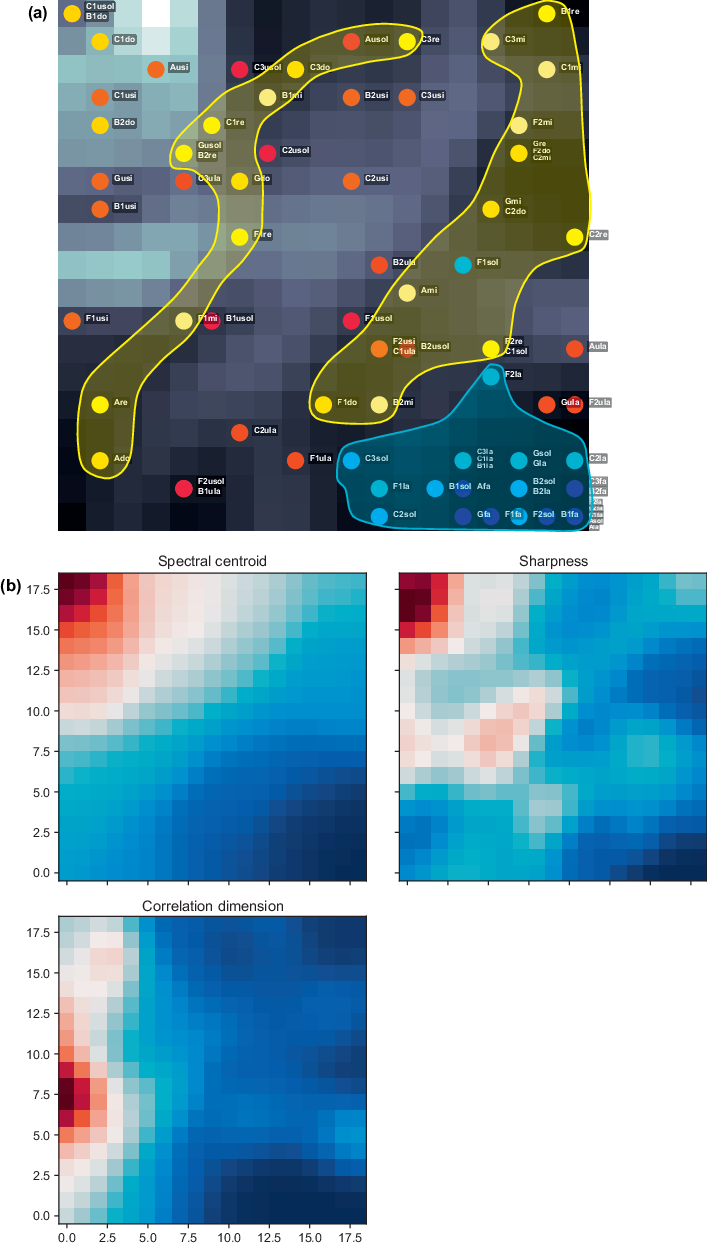}
\end{minipage}
\hspace{4mm}
\begin{minipage}[t]{0.23\textwidth}
\vspace{0pt}
\centering
\includegraphics[width=\linewidth]{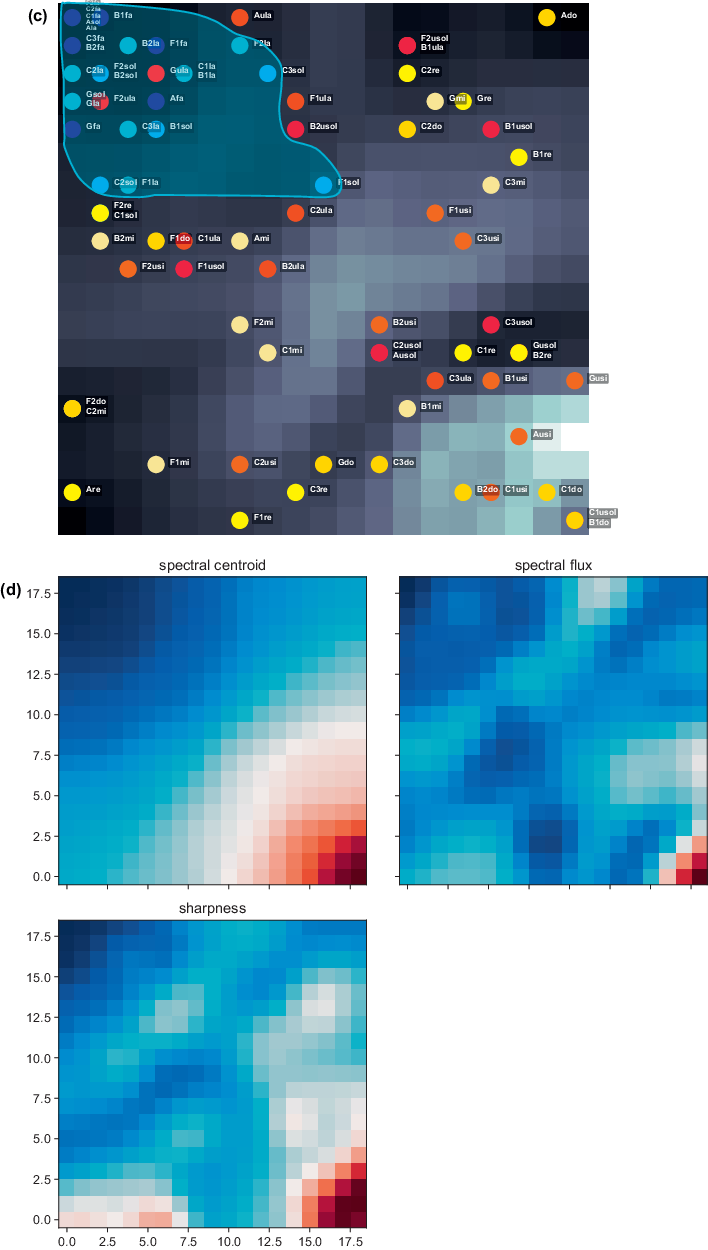}

\vspace{1.5mm}

\includegraphics[width=\linewidth]{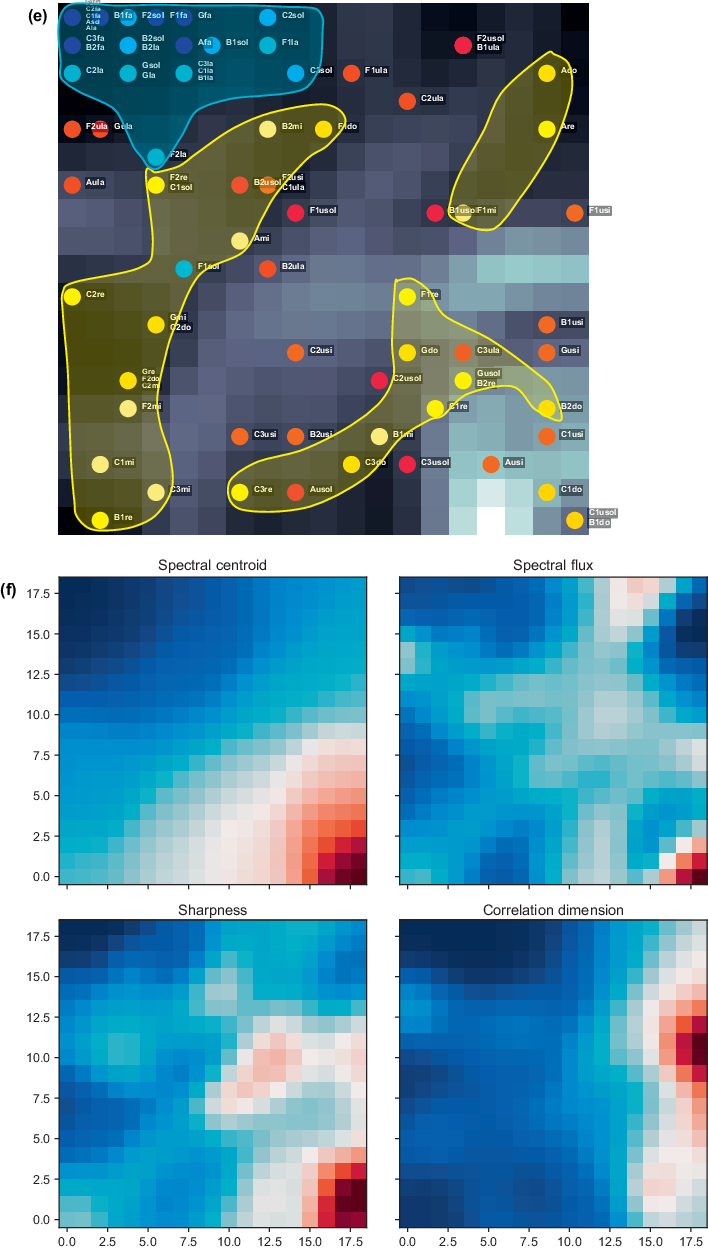}
\end{minipage}

\vspace{-1mm}

\caption{\label{fig:FIG6}
Multi-feature SOM maps with neighbouring neuron similarity as background color.
The large left panel shows the SOM and feature maps obtained using spectral centroid,
sharpness, and correlation dimension. The two smaller panels on the right show the SOM
and feature maps obtained using spectral centroid, spectral flux, and sharpness, and using
spectral centroid, sharpness, spectral flux, and correlation dimension, respectively.
}
\end{figure*}

First, a feature vector consisting of the spectral centroid, sharpness, and fractal correlation dimensions is used, and the results are displayed in Fig.~\ref{fig:FIG6}(a). The combination results in three clusters: a cluster region of the treble sounds in the lower right corner with darker background color; a cluster of midrange tones dispersed in the upper right region with a slightly lighter background color in the middle; and a cluster of bass sounds dispersed in the lower left region with a somewhat lighter background color. Thus, compared with the single-feature maps, the treble sounds are clustered better with the multifeature maps, forming a dense dark area. The midrange tones are more dispersed than the high-pitched tones are, forming two slightly lighter-colored clustered areas, as shown by the yellow curve region in the figure. The bass area is also relatively more dispersed, scattered primarily below the lighter-colored clustered boundary in the upper left corner, down to the darker area at the bottom. Therefore, compared with the single-feature maps, the inclusion of this feature vector improves the clustering of high-pitched tones.

The feature maps in which the three features are used to determine the background colors are shown in Fig.~\ref{fig:FIG6}(b). The spectral centroid and sharpness plots are similar and nearly orthogonal to the fractal correlation dimension. The spectral centroid feature shows the most continuous gradient, indicating that this feature is the most prominent feature among the three features.

The sharpness feature is also important, with the addition of this feature improving the clustering results, with complex structures observed across various regions. The fractal dimension adds another component, enabling better differentiation of the clusters.

The high-frequency area of the spectral centroid feature is located in the upper left corner, whereas the fractal correlation dimension feature is strongest in the middle and lower parts (vertical coordinates 7--9). Two intersecting ridges are roughly distributed across the spectral centroid and fractal correlation dimension features, as shown in Fig. 6(a). The treble area of the \emph{hulusi} is distributed in the lower right region, where the two ridges intersect, also known as the dark weak distribution area. Most of the midrange area corresponds to the weak area of the fractal correlation dimension feature, specifically the upper right corner of Fig.~\ref{fig:FIG6}(a). There is no discernible pattern in the distribution of the bass area.

The results of the spectral centroid, spectral flux, and sharpness feature combination are shown in Fig.~\ref{fig:FIG6}(c). Only the treble region is clustered in the upper left corner, and the midrange and bass tones are not clustered. The midrange tones spread to the lower right region as the pitch decreases, with a light-colored clustering boundary area in the lower right corner. As the pitch decreases, the bass tones spread in the opposite direction, toward the upper left corner. A few do and usi sounds can be observed on the light cluster boundaries, indicating that their spectral flux combination characteristics differ significantly from those of other sounds. The original map with the spectral centroid, spectral flux, and sharpness parameters used to determine the background colors is shown in Fig.~\ref{fig:FIG6}(d). The three characteristic distributions are not mutually orthogonal. The spectral flux features are distributed primarily in the lower right region, weakens toward the upper left region, and then strengthens again directly below this region (horizontal coordinates 8--10). The sharpness features are distributed mainly in the lower right region, with secondary clustering areas in the lower left (horizontal coordinates 4--6) and upper right (vertical coordinates 0- 2) regions. The corresponding characteristic responses are shown in Fig.~\ref{fig:FIG6}(c), where the treble tones are ideally in the weak region of the three parameters in the upper left corner. While the midrange and bass tones are dispersed, they are essentially clustered in the weak areas of the sharpness and spectral flux features but not in the weak regions of the spectral centroid feature.

Next, four parameters were combined: the spectral center of mass, spectral flux, sharpness, and fractal correlation dimension (Fig.~\ref{fig:FIG6}(e)). The treble tones remain perfectly clustered in the darker-colored region in the upper left corner. The midrange tones are divided across three slightly lighter-colored subclusters in the center (vertical coordinates 6--10), upper right (horizontal coordinates 6--17), and upper (vertical coordinates 13--17) regions, as shown by the yellow region in the figure. However, the bass sounds could not be clustered and were scattered throughout the figure. Individual \emph{do} and \emph{usi} sounds are observed in the lighter-colored region on the lower right cluster boundary, corresponding to the homophonic pitches in the previous feature combination plot. The spectral centroid and fractal correlation dimension features are nearly orthogonal in the background image of the four parameters (Fig.~\ref{fig:FIG6}(f)).

In contrast, the spectral flux and sharpness features are clustered into more than two main areas. As a result, in addition to the most substantial area (vertical coordinates 13--15), Fig.~\ref{fig:FIG6}(e) shows two nearly parallel ridges running from the top (horizontal coordinates 7--9) to the lower right (horizontal coordinates 13--14) region and from the upper left (vertical coordinates 14--15) to the bottom (horizontal coordinates 7--8) region.

The high-pitch area of the \emph{hulusi} clusters below the second ridge, which represents the weakest of the four parameters (the dark position in the lower left corner).

The midrange tones are concentrated outside the two ridges in the weak areas of the spectral flux, sharpness and fractal correlation dimension features. The bass tones are clustered across highly irregular regions.

These findings show that including the spectral spread, a feature that is poorly clustered when used alone, is not considerably beneficial for improving the clustering performance when it is included in feature vector combinations.

\subsection{\label{subsec:3:3} Calculating the Cluster Quality}
To quantitatively verify the clustering of different features of each pitch in the above SOMs, a cluster quality value $C^{p,f}$ is calculated as the fraction between $S^{p,f}$, the mean of the distances $d^{p,f}(i,j)$ of feature f between all notes i and j of the same pitch p over $O^{p,f}$, and the mean of the distances $d^{p,f}(i,j)$ between all notes i with pitch p to all notes j of all other pitches.

Therefore, for each pitch p and feature f, the mean distance $d^{p,f}(i,j)$ between all notes i of pitch p and all notes j of the same pitch p is as follows:

\begin{equation}
	S^{p,f} = \frac{\sum_{i,j}  d^{p,f}(i,j)}{N^{\in p} \times (N^{\in p} - 1)} \  \forall i \in p \ \&\ \forall j \in p \  \& \ i \neq j \ .
\end{equation}

Furthermore, the mean distance $d^{p,f}(i,j)$ between all notes i of pitch p and all notes j of all other pitches is as follows:

\begin{equation}
	O^{p,f} = \frac{\sum_{i,j}  d^{p,f}(i,j)}{N^{\in p} \times M^{\notin p}} \  \forall i \in p \ \&\ \forall j \notin p.
\end{equation}

Here, $N^{\in p}$ is the number of notes with pitch p, and $M^{\notin p}$ is the number of notes with a pitch other than p.

The pitches p of all the \emph{hulusi} tones are arranged from p = 0 (lowest) to p = 9 (highest). The distances $d^{p,f}(i,j)$ are calculated as the correlation between the normalized feature vectors of notes i and j.

The fraction between the mean distances of all notes with the same pitch $S^{p,f}$ over all notes with other pitches $O^{p,f}$ is the cluster quality index, which is calculated as follows:

\begin{equation}
	C^{p,f} = \frac{S^{p,f}}{O^{p,f}} \ .
\end{equation}

If the notes of the same pitch p for a feature f are clustered on the SOM, $S^{p,f}$ is small and $O^{p,f}$ is large, and vice versa. Therefore, small $C^{p,f}$ values indicate successful clustering, whereas $C^{p,f} \geq 1$ indicates weak or no clustering.

Furthermore, the mean over all the features is calculated as follows:

\begin{equation}
	C^p = \frac{\sum_{i=1}^{N_f} C^{p,f}}{N_f} \ ,
\end{equation}

where $N_f = 7$ is the number of timbre features.

\begin{figure}
    \centering
    \includegraphics[height=0.7\columnwidth]{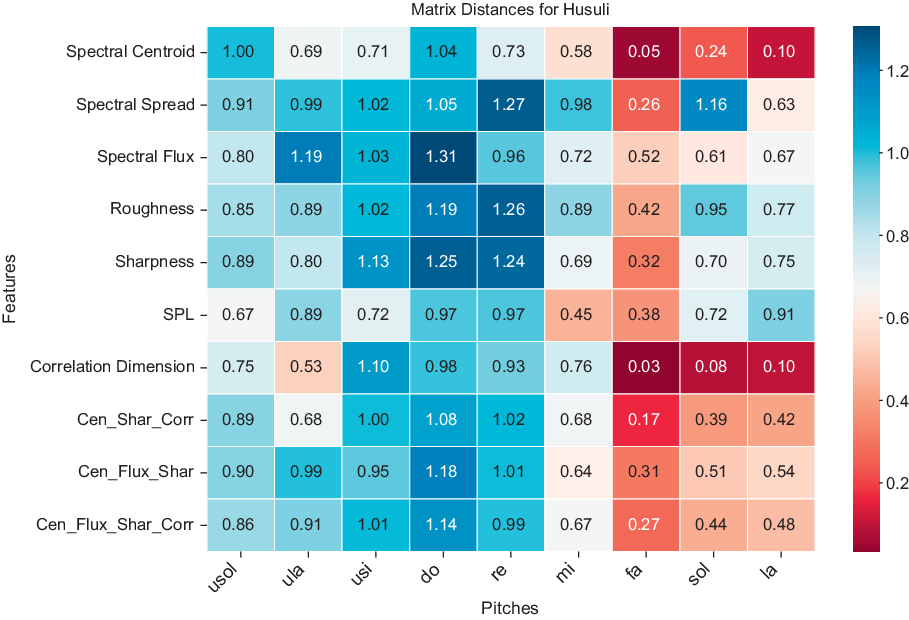}
    \caption{\label{fig: FIG7}{Distances of \emph{hulusi} pitches, indicating the ability to form clusters for the single features and feature combinations. Pitches are arranged from low to high; purple indicates close distances and good clustering, whereas green indicates far distances and poor clustering. The last three rows of the vertical axis are the features discussed in Section. 3.2.}}
\end{figure}

Fig.~\ref{fig: FIG7} shows the distances between all the used features of each pitch for the nine \emph{hulusi} instruments. The three bottom rows show the matrix distance of the feature combination discussed in Section 3.2. Dark purple corresponds to smaller distances, while dark green indicates larger distances. The values for \emph{fa, sol,} and \emph{la} in the high-pitched range are much less than one, indicating that their distances are very small, indicating nearly perfect clustering.

The most apparent features are the fractal correlation dimension and the spectral centroid, which are consistent with the results of the Kohonen SOMs. On the other hand, the distances of usi, do, and re, which are midrange and base tones, are large, and the notes do not form clusters. Finally, the \emph{usol, ula,} and \emph{mi} sounds are between the above two cases and form loose clusters, which is consistent with the visual observations. Compared with the use of a single feature, the use of the feature combination discussed in section 3.2 can result in more obvious clusters (except \emph{usi, do,} and \emph{re}).

\begin{figure}
    \centering
    \includegraphics[height=0.55\columnwidth]{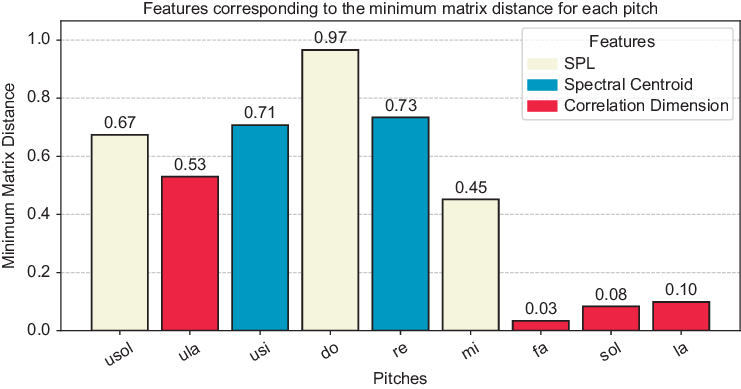}
    \caption{\label{fig: FIG8}{Timbre features corresponding to the minimum distances for all pitches from Fig. 7. The fractal correlation dimension is the best feature for clustering \emph{hulusi} tones.}}
\end{figure}

The features and values of the shortest distances for each pitch are summarized in Fig.~\ref{fig: FIG8}.

The shortest distances for all the pitches mainly involve the SPL, spectral centroid, and fractal correlation dimension features.
The feature corresponding to the minimum distances of the high-pitched \emph{fa, sol,} and \emph{la} sounds is the fractal correlation dimension, whose values are much smaller than those of all the other pitches. Therefore, the fractal correlation dimension is the best feature for clustering \emph{hulusi} tones.

Determining the acoustic reason why the notes \emph{fa, sol,} and \emph{la} clearly clustered. Better than the other pitches and why the fractal dimension is the best-fit feature is beyond the scope of the paper. Nevertheless, the uppermost fingerhole on the front side of the main tube is left open only when the three notes \emph{fa, sol,} and \emph{la} are played. Therefore, compared with the other pitches, the higher pitches need to be played with less pressure. Interestingly, the fractal correlation dimension feature enables the best clustering of these pitches. This feature is associated mainly with chaoticity in the initial transient, which appears to be substantially reduced with this playing technique. Additionally, the reduced brightness might be caused by the reduced playing pressure.

\section{\label{sec:5}Discussion}
SOMs provide a comprehensive view of the \emph{hulusi's} timbral characteristics using psychoacoustic training features. We investigated the variation of the timbral characteristics of different \emph{hulusi} modes, inspired by recent studies in which the SOMs for all feature orders were improved and an SOM-acoustic SOMson based on a four-dimensional feature space \citep{Linke2024} was proposed. In the analysis, we used nine \emph{hulusi} made by three instrument manufacturers in Yunnan Province, played in an anechoic chamber, recorded with a dummy head, psychoacoustically analyzed, and trained with the \emph{COMSAR} framework.

In the measurement results of the Yunnan hulusi, the self-organized clustering regions of the spectral spread, roughness, and SPL are not obvious, and the distribution of the pitch is relatively chaotic. As a result, in this paper, the timbre of the Yunnan hulusi is analyzed using four features: the spectral centroid, spectral flux, sharpness, and fractal correlation dimension.

Using single-feature SOMs, the treble and midrange regions are best clustered using the sharpness features. The treble region is only clustered using the spectral centroid, sharpness and fractal correlation dimension, and there is no apparent clustering for the spectral flux feature. The single-feature clustering results are not obvious, except for sharpness, spectral centroid and fractal correlation dimension. According to the values of the matrix distances in Fig.~\ref{fig:FIG7}, \emph{mi, fa, sol,} and \emph{la} can be clustered according to the sharpness, and \emph{fa, sol,} and \emph{la} can be clustered according to the spectral centroid and fractal correlation dimensions. The \emph{ula, usi, re,} and \emph{mi} values for the spectral centroid feature and the \emph{usol, ula,} and \emph{mi} values for fractal correlation dimension feature are relatively big; they(refer to \emph{ula, usi, re} and \emph{mi}) do not form apparent clusters in the SOMs, so a combination of plots and numerical values is needed to obtain accurate conclusions.

In the multifeature SOMs, the clustering of the \emph{hulusi's} pitches is much more apparent. In Figs.~\ref{fig:FIG6}(a) and (b), the mid-pitch region clusters around low fractal correlation dimension values, implying that the initial transients are small, which may be related to the midrange's proximity to the reed's eigenfrequency. The treble region is clustered in both low-value regions of the fractal correlation dimension, and the corresponding spectral centroid clustering region is also in the low-value region, indicating that the \emph{hulusi's} treble tones have small initial transients and fewer overtones.

In Figs.~\ref{fig:FIG6}(c) and (d), only the treble sounds is clustered in the low-valued spectral centroid region. In contrast, neither the alto nor the bass sounds are clustered in the low spectral center of mass region, most likely because of the fewer overtones of these sounds.

The results in Fig.~\ref{fig:FIG6}(e) and (f) are more consistent with the expected pattern than the results in (c) and (d). According to (f), the spectral centroid feature is nearly perfectly orthogonal to the fractal correlation dimension feature.

In contrast, the two inconspicuous ridges are roughly orthogonal to the fractal correlation dimension in both the sharpness and spectral flux SOMs. As a result of the chaotic nature of the initial transient, which is expressed via the fractal correlation dimension, the treble, midrange, and bass tones of the \emph{hulusi} can be clustered, thereby enabling the \emph{hulusi} timbres to be distinguished.

The multifeature combination results in Fig.~\ref{fig: FIG7} are consistent with the above characteristics. In addition, the values of ula and mi are relatively small. Furthermore, the results in Fig.~\ref{fig:FIG6}(a), (c) and (e) reveal that ula clusters next to a slightly lighter region, whereas mi does not cluster in the SOM despite its small value. In contrast, \emph{usi, do,} and \emph{re} have large values and do not form clusters. This also explains why \emph{do} and \emph{usi,} as discussed in Section 3.2, always appear on the light-colored boundary line in the SOM.

The above findings significantly improve our understanding of the clustering of the spectral centroid and fractal correlation dimension features. For example, the \emph{COMSAR} music AI framework can cluster \emph{hulusi} timbre characteristics on the basis of the fractal correlation dimension and spectral centroid features. This also serves as the foundation for future work on physical analysis and modeling of \emph{hulusi} reeds.

The next step is to understand the mechanical and acoustical reasons for these timbre features, which can be performed through ongoing measurements of the reed vibrations by considering the coupling between the reed and tube eigenfrequencies, which will be performed in future work.

\section*{Author Declarations}
\subsection*{Conflict of Interest}
The authors have no conflicts to disclose.

\subsection*{Data Availability}

The recordings that were used to train the SOMs are not publicly available but can be requested from the author via email.

\vspace{1cm}
\textbf{Literature}

\bibliography{citation-1}

@article{Braasch,
author = {Braasch, Jonas},
year = {2023},
month = {01},
pages = {14},
title = {Free Reeds: An Intertwined Tale of Asian and Western Musical Instruments},
volume = {19},
journal = {Acoustics Today},
doi = {10.1121/AT.2023.19.4.14}}

@article{cottingham2011,
  title={Acoustics of free-reed instruments},
  author={Cottingham, James},
  journal={Physics Today},
  volume={64},
  number={3},
  pages={44--48},
  year={2011},
  publisher={AIP Publishing}}

@InProceedings{Mcfee2015,
  author    = { {B}rian {M}c{F}ee and {C}olin {R}affel and {D}awen {L}iang and {D}aniel {P}.{W}. {E}llis and {M}att {M}c{V}icar and {E}ric {B}attenberg and {O}riol {N}ieto },
  title     = { librosa: {A}udio and {M}usic {S}ignal {A}nalysis in {P}ython },
  booktitle = { {P}roceedings of the 14th {P}ython in {S}cience {C}onference },
  pages     = { 18 - 24 },
  year      = { 2015 },
  editor    = { {K}athryn {H}uff and {J}ames {B}ergstra },
  doi       = { 10.25080/Majora-7b98e3ed-003 }}

@article{Fletcher1978,
    author = {Fletcher, N. H.},
    title = "{Mode locking in nonlinearly excited inharmonic musical oscillators}",
    journal = {The Journal of the Acoustical Society of America},
    volume = {64},
    number = {6},
    pages = {1566-1569},
    year = {1978},
    month = {12},
    issn = {0001-4966},
    doi = {10.1121/1.382139},
    url = {https://doi.org/10.1121/1.382139},
    eprint = {https://pubs.aip.org/asa/jasa/article-pdf/64/6/1566/12107984/1566\_1\_online.pdf},}

@article{Bader1,
  author       = {Rolf Bader and
                  Michael Bla{\ss} and
                  Jonas Franke},
  title        = {Computational timbre and tonal system similarity analysis of the music
                  of Northern Myanmar-based Kachin compared to Xinjiang-based Uyghur
                  ethnic groups},
  journal      = {CoRR},
  volume       = {abs/2103.08203},
  year         = {2021},
  url          = {https://arxiv.org/abs/2103.08203},
  eprinttype    = {arXiv},
  eprint       = {2103.08203},
  bibsource    = {dblp computer science bibliography, https://dblp.org}}

@book{Kohonen1997, 
      title={Self-Organizing Maps}, 
      ISBN={9783642979668}, 
      ISSN={0720-678X}, 
      url={http://dx.doi.org/10.1007/978-3-642-97966-8}, 
      DOI={10.1007/978-3-642-97966-8}, 
      journal={Springer Series in Information Sciences}, 
      publisher={Springer Berlin Heidelberg}, 
      author={Kohonen, Teuvo}, 
      year={1997} }

@article{Linke2024,
  title={SOMson--Sonification of Multidimensional Data in Kohonen Maps},
  author={Linke, Simon and Ziemer, Tim},
  journal={arXiv preprint arXiv:2404.00016},
  year={2024}}

@article{Bader2,
author={Rolf Bader and Axel Zielke and Jonas Franke},
journal={Journal of the audio engineering society},
title={Timbre-based machine learning of clustering chinese and western hip hop music},
year={2021},
number={10473},
month={may} }

@misc{BlassWeb,
    author = "Bla{\ss}, Michael and Bader, Rolf",
    title = "COMSAR framework",
    howpublished = "Website",
    year = {2019},
    note = {\url{https://gitlab.rrz.uni-hamburg.de/bal7668/comsar}}
}

@article{Blass2019,
  title={Content-based music retrieval and visualization system for ethnomusicological music archives},
  author={Bla{\ss}, Michael and Bader, Rolf},
  journal={Computational Phonogram Archiving},
  pages={145--173},
  year={2019},
  publisher={Springer}}

@book{Zwicker2013,
  title={Psychoacoustics: Facts and models},
  author={Zwicker, Eberhard and Fastl, Hugo},
  volume={22},
  pages={257--264},
  year={2013},
  publisher={Springer Science \& Business Media}}

@book{Fastl2007,
  title={Psychoacoustics: Facts and models},
  author={Fastl, Hugo and Zwicker, Eberhard},
  volume={3rd Edition},
  pages={239--243},
  year={2007},
  publisher={Springer Berlin Heidelberg}}

@book{Rossing2007,
  title={Springer handbook of acoustics},
  author={Rossing, Thomas},
  year={2007},
  publisher={Springer Science \& Business Media}}

@book{Bader2013,
  title={Nonlinearities and synchronization in musical acoustics and music psychology},
  author={Bader, Rolf},
  volume={2},
  pages={94--100},
  year={2013},
  publisher={Springer Science \& Business Media}}

@article{Mcadams2013,
  title={Musical timbre perception},
  author={McAdams, Stephen},
  journal={The psychology of music},
  volume={3},
  year={2013}}

@article{Alluri2010,
  title={Exploring perceptual and acoustical correlates of polyphonic timbre},
  author={Alluri, Vinoo and Toiviainen, Petri},
  journal={Music Perception},
  volume={27},
  number={3},
  pages={223--242},
  year={2010},
  publisher={University of California Press USA}}

@article{Yang2017Instrument,
  title={A review of research on musical instruments of ethnic minorities in Yunnan since the 1980s},
  author={Yang, Chen},
  journal={Journal of Xinhai Conservatory of Music},
  number={4},
  pages={10},
  year={2017}}

@article{Lou2014,
  title={Stimulate interest, practice diligently, and gain something from it:  experience of introducing Hulusi into the classroom},
  author={Jiaqi Lou},
  journal={Music Space},
  number={5},
  pages={1},
  year={2014}}

@article{Liu2022,
  title={Research on Hulusi teaching method under the spirit of inheriting Chinese aesthetics},
  author={Liu, Li},
  journal={China National Exhibition},
  number={21},
  pages={90-93},
  year={2022}}

@article{Liu2014,
  title={The grand event of hulusi art — the first hulusi Art Festival in Kunming in 2014},
  author={Liu, Mo},
  journal={National Music(China)},
  number={05},
  pages={90},
  year={2014}}

@article{Zhang2023,
  title={From the beginning of the year to the end of the Lianghe cultural travel high non-stop},
  author={Zhang, Fan},
  journal={Dehong Unity Newspaper},
  number={22.12},
  pages={1},
  year={2023}}

@article{Dieckman2006,
author = {DIECKMAN, ERIC},
year = {2006},
month = {11},
pages = {},
title = {Input impedance of asian free-reed mouth organs},
volume = {120},
journal = {The Journal of the Acoustical Society of America},
doi = {10.1121/1.4787404}
}

@article{Cottingham2000,
author = {Cottingham, James},
year = {2000},
month = {05},
pages = {},
title = {Acoustics of a symmetric free reed coupled to a pipe resonator},
volume = {107},
journal = {Journal of The Acoustical Society of America - J ACOUST SOC AMER},
doi = {10.1121/1.428770}
}

@article{Martinez2025,
author = {Martinez, Cristhiam and Bader, Rolf},
year = {2025},
month = {03},
pages = {},
title = {Analysis of nonlinear behavior of viscoelastic damping in musical membranes using physics-informed self-organizing maps},
volume = {3},
journal = {APL Machine Learning},
doi = {10.1063/5.0242985}
}

\end{document}